\documentclass[aps,prl,twocolumn,groupedaddress]{revtex4}

\usepackage{graphicx}
\usepackage{dcolumn}

\begin{document}


\title{Coupled atomic motion and spin-glass transition in FeAl$_{2}$}


\author{Yang Li}
\author{F. G. Vagizov}
\author{V. Goruganti}
\author{Joseph H. Ross, Jr.}
\affiliation{Department of Physics, Texas A\&M University, College
Station, TX 77843-4242}

\author{Zuxiong Xu and Guohui Cao}
\affiliation{Department of Physics, University of Science and
Technology Beijing, Beijing 100083, China}

\author{Cheng Dong}
\affiliation{National Laboratory for Superconductivity, Institute
of Physics, Beijing 100080, China}

\author{Zhaosheng Feng}
\affiliation{Department of Mathematics, University of Texas-Pan
American, Edinburg, TX 78541}


\date{\today}

\begin{abstract}
We have used $^{57}$Fe M\"{o}ssbauer spectroscopy and x-ray
diffraction to study the magnetic and 
vibrational properties of FeAl$_{2}$. FeAl$_{2}$ is an
ordered intermetallic with a large Fe local moment, and a complex
crystal structure with site-occupation disorder on some sites. This material
exhibits spin-glass freezing below 35 K. 
From the $^{57}$Fe recoilless fraction, we
find that there is a vibrational mode which freezes out
concurrently with the spin freezing. X-ray powder diffraction measurements
confirm this result, indicating an anomalous change in the Debye-Waller
factor at temperatures below the spin-freezing temperature. 
\end{abstract}

\maketitle


Coupling of structural distortions to a magnetic transition can
lead to a number of interesting physical phenomena. For instance,
in the manganite La$_{1-x}$Ca$_{x}$MnO$_{3}$, a Jahn-Teller
distortion coupled to the ferromagnetic transition significantly
enhances the magnetoresistance \cite{Asam95nat,Rami97JPC}, leading
to the colossal magnetoresistive effect. Furthermore, in a range
of manganite and cuprate materials, the competition between
magnetism and structural distortions contributes to the
development of stripes and related textures, which has led to a
wealth of new physics \cite{Renn02nat,Beck02PRL,Math03pto}, and
plays an important role in high-$T_{c}$ superconductivity.
The interplay of structural distortion and a spin glass
transition has also been identified in the highly frustrated
pyrochlore lattice: the pyrochlore Y$_{2}$Mo$_{2}$O$_{7}$ exhibits
a lattice distortion driven by a spin-glass transition, which
allows the magnetic frustration to be reduced \cite{Kere01PRL}.  In
this letter, we describe M\"{o}ssbauer and x-ray measurements indicating
that FeAl$_{2}$ has a soft mode which is coupled to its spin-freezing
transition. This soft mode becomes activated above the spin freezing
temperature ($T_{f}$), thus is coupled to the magnetic excitations.

\begin{figure}
\includegraphics[width=\columnwidth]{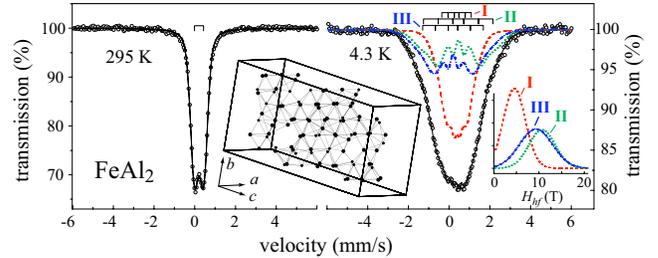}%
\caption{\label{fig:figA}M\"{o}ssbauer spectra for FeAl$_{2}$ at 295 K
and 4.3 K. Inset at right: magnetic hyperfine field distribution 
for 4.3 K lineshape fit.  Center: A portion of FeAl$_{2}$
structure (frame=block of 3$\times$2$\times$3 unit cells) viewed perpendicular to 
nearly-close-packed planes.}
\end{figure}

FeAl$_{2}$
is an ordered intermetallic with a complex structure that can be described
as containing layers which are roughly close-packed \cite{Corb73acr}. 
Inset within Fig.~\ref{fig:figA} is a view of this structure perpendicular to these layers.
There are ten Al and five Fe sites per unit cell, with three additional sites having mixed Fe-Al 
occupation, plus one vacancy site per cell, forming channels along $a$. The vacancies 
presumably help 
to adjust the electron count in order to provide energy stabilization via
a pseudogap at the Fermi surface \cite{Kraj02JPC}; hence the observed
semimetallic behavior \cite{Lue03JPC}. The Fe-Fe coordination number is
small, however the effective magnetic moment per Fe atom (2.5 $\mu_{B}$) is
large \cite{Lue01PRB}, and has been shown to be a stable local 
moment on Fe \cite{chi05prb}.

The magnetic behavior of FeAl$_{2}$ is dominated by a spin freezing
transition at $T_{f}$ = 35 K \cite{Lue01PRB}, below which the magnetization exhibits
the typical frequency dependence and hysteretic behavior of a spin
glass. The coupling between Fe spins is dominated by 
superexchange through Al orbitals, which is antiferromagnetic
in sign. The complex structure and mixed occupancy of some sites 
contribute to the magnetic frustration, and hence the spin glass
configuration. NMR measurements \cite{chi05prb} have shown that the
spin fluctuations at temperatures well above $T_{f}$ are dominated by 
non-thermal flip-flop processes between super-exchange-coupled Fe moments. 
Here we show, using M\"{o}ssbauer spectroscopy and
x-ray diffraction, that these fluctuations are tied to a soft
vibrational mode that persists to low temperatures, and which is
strongly coupled to the magnetic behavior.

Samples for this study were prepared by arc melting Fe and Al, and subsequent
solid-state reaction, to yield polycrystalline ingots. Rietveld analysis 
of the powder x-ray diffraction pattern for this sample has been discussed
previously \cite{chi05prb}; no additional phases could be
detected. M\"{o}ssbauer spectra were obtained
using a $^{57}$Co source in a Rh matrix. The sample was ground and
sieved to less than 20$\mu$m. Shift and
velocity calibrations are referred to $\alpha$-Fe at room
temperature.

\begin{figure}
\includegraphics[width=\columnwidth]{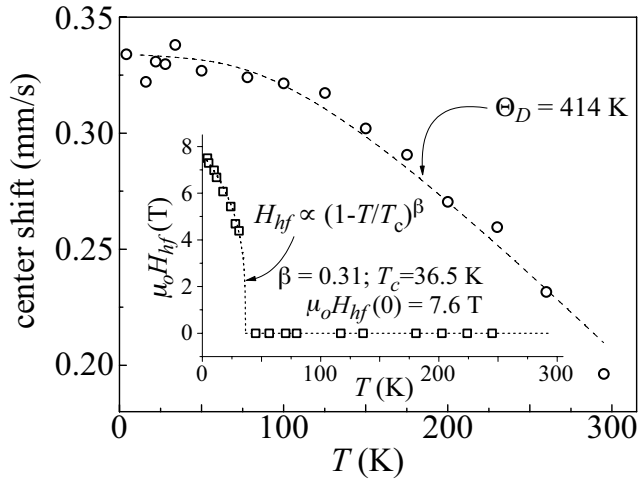}
\caption{\label{fig:figD}Weighted average M\"{o}ssbauer center shift, obtained 
from fitted curves. Dashed curve: fit to Debye vibrational model, with $\Theta_{D}$ = 414
K.  Inset: weighted average hyperfine fields.}
\end{figure}

M\"{o}ssbauer spectra taken at 4.3 K and 295 K are shown in Fig.~\ref{fig:figA}. A
large increase in linewidth commences at $T_{f}$, which
can be attributed to the development of quasi-static
magnetic hyperfine fields ($H_{hf}$) at low temperatures \cite{chi05prb}.
Absorbtion lines include a superposition of
five inequivalent Fe sites, plus the three mixed sites,
within the triclinic FeAl$_{2}$ structure (space
group P1; see inset of Fig.~\ref{fig:figA}). 
To fit the hyperfine field distributions, here we have used a Voigt-based fit according to 
a method described previously \cite{Xu93CSB,Rau91NIB}. 
The fit used three broadened multiplets to approximate the distribution of quadrupole and magnetic
hyperfine fields, with a low-temperature distribution of $H_{hf}$ shown as an inset to
Fig.~\ref{fig:figA}. 
The resulting weighted average $H_{hf}$, shown as the inset to Fig.~\ref{fig:figD},
develops sharply below $T_{f}$, quite similar to the previous result  \cite{chi05prb}, but
with slightly larger average values of $H_{hf}$ obtained with this model.

\begin{figure}
\includegraphics[width=\columnwidth]{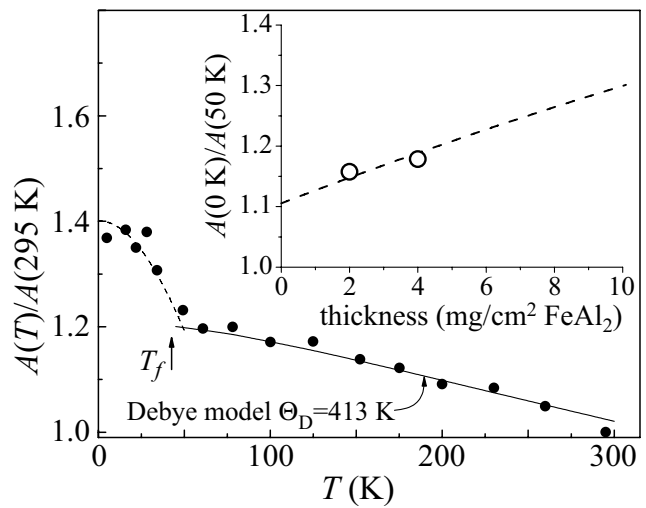}
\caption{\label{fig:figB}FeAl$_{2}$ M\"{o}ssbauer spectral areas. 
Solid curve: Debye fit with $\Theta$$_{D}$ =
413 K. Dashed curve: phenomenological fit to low-temperature
anomaly, described in text. Inset: anomalous area ratio vs.
sample density, with
theoretical fit.}
\end{figure}

The temperature dependence of the relative M\"{o}ssbauer
spectral areas is shown in Fig.~\ref{fig:figB}. The data
were obtained for a sample containing 4 mg/cm$^{2}$
FeAl$_{2}$. The area was
obtained directly from the measured spectra by means of Simpson 
integration using a
fitted-background subtraction, so as to be independent of the spectral
fitting (Fig.~\ref{fig:figA}). The area is related to the
recoilless fraction ($f$) \cite{vertes79}; above $T_{f}$
the gradual reduction follows the normal trend, due to increased atomic motion 
because of the population of phonon modes. Further changes in $f$ at low
temperatures imply the freezeout of an additional lattice vibrational mode, or perhaps
a localized mode involving loosely-bound atoms \cite{vog76prl,pet80prl,pau81jssc}.

M\"{o}ssbauer areas are related to nonlinear absorbtion 
in thick
samples, as well as to $f$. To investigate the thickness
dependence, we prepared a lower density sample. 
The inset of Fig.~\ref{fig:figB} shows area ratios between $T$ = 0 K
and 50 K, encompassing the magnetic changes. 
$T$ = 0 values were obtained by fitting to a function of the form,
exp($- T^{2}/T_{o}^{2}$) (dashed curve in the main plot of
Fig.~\ref{fig:figB}). The inset plot shows a fit to the change vs. thickness, based on
the known behavior of M\"{o}ssbauer spectral areas \cite{vertes79}. 
This calculation assumed a change
from a doublet to a sextet; although in our case the individual lines are not
resolved, this assumption agrees well with the observed
linewidth change. The only adjustable parameter for the inset curve was
an additive constant, found equal to 0.1 (seen by the intercept), which
represents  the change in $f$ once the thickness effect is removed.
Thus the observed changes in area are due in large part to a change in
vibrational properties at low-temperatures.

For the case of nuclear motion, $f$ is reduced due to
dynamical phase shifts of the radiation, averaged 
over the radiative lifetime. Harmonic motion yields to good
approximation \cite{Dickson1}
$f$ = exp($-k^{2}\langle x^{2}\rangle$) $\equiv$ exp(-2$W$) where 2$W$ is
Lamb-M\"{o}ssbauer factor, $k$
is the $\gamma$-ray wavevector and $\langle x^{2}\rangle$ the
mean square linear displacement. Using a Debye phonon
model one finds \cite{Dickson1}
\begin{equation}
    2W = \frac{6E_{R}}{k_{B}\Theta_{D}}\left[\frac{1}{4}+\frac{T^{2}}{\Theta_{D}^{2}}\int_{0}^{\Theta_{D}/T}\frac{\xi d \xi}{e^{\xi}-1}\right]
\label{eq:one},
\end{equation}
where $E_{R}$ and $\Theta_{D}$ are the recoil energy
and Debye temperature respectively. A fit of the high-temperature
data gives $\Theta_{D}$ = 413 K, shown as a solid curve in Fig.~\ref{fig:figB}. 
This is comparable to that of other intermetallic compounds [for example
Zr(Fe,Al)$_{2}$ has $\Theta_{D}$ = 420 K \cite{Isra97JAC}]. The increased
area below $T_{f}$ is a departure from the Debye model. 
Lattice softening at a structural phase transition typically gives an
enhancement of $f$ only in the critical region near
$T_{c}$ \cite{Dickson1}. In manganites and
related colossal magnetoresistive materials, $f$ increases
at temperatures somewhat below $T_{c}$ \cite{nat02prb},
attributed to fluctuations confined to nanoscale clusters
associated with polarons. For FeAl$_{2}$, the reduction in
$f$ from $T$ = 0 to $T_{f}$ corresponds to
the activation of a vibrational mode with
$\sqrt{\langle x^{2}\rangle}$ = 0.0044 nm for Fe,
calculated according to $f$ = exp($-k^{2}\langle x^{2}\rangle$).

\begin{figure}
\includegraphics[width=\columnwidth]{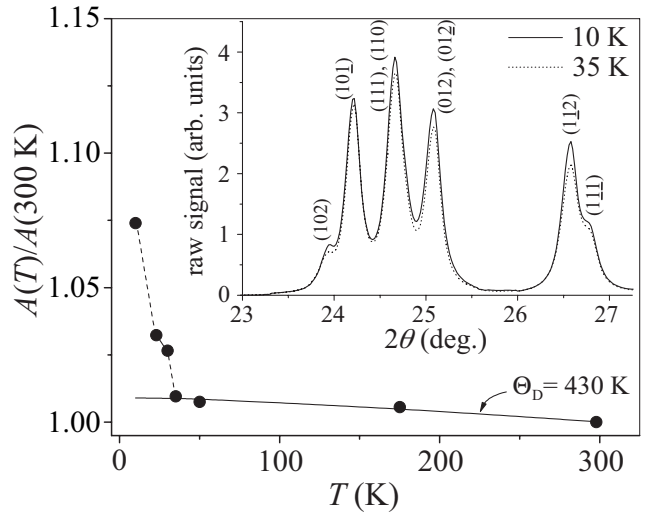}
\caption{\label{fig:figC}Relative x-ray diffraction
intensities for FeAl$_{2}$ vs. $T$. Solid curve:
Debye fit with $\Theta_{D}$ = 430 K. Dashed curve connects the low-$T$ points,
showing an anomalous increas. Inset: x-ray diffraction patterns at 10 and 35 K.}
\end{figure}

Low temperature x-ray diffraction measurements also support the
presence of a low-temperature vibration anomaly. We examined the
diffraction profile at temperatures between 10 and 300 K, over a restricted angular range of
19$^{\circ }$ to 27$^{\circ }$, which contains some of the characteristic
peaks. (A full powder spectrum
was shown in Ref. \cite{chi05prb}.) The relative peak
intensities were obtained by means of Simpson integration using
a fitted-background subtraction \cite{don99jac}. From the
diffraction profiles we see no evidence of any
structural transition down to 10 K (the instrument limitation), though
the limited angular range precludes a detailed structural analysis.
The inset to Fig.~\ref{fig:figC} compares data from 10 K and 35 K, for the region containing
the most intense peaks, showing the amplitude increase with little or no 
peak shift.
A striking feature is that the
relative intensities exhibit a rapid increase below 35 K, similar
to the M\"{o}ssbauer results. This implies an
anomalous stiffening as the temperature decreases below
$T_{f}$. 

The main plot of Fig.~\ref{fig:figC} shows the evolution of the peak areas.
The solid curve is a fit of the high-temperature data 
to exp(-2$M$), where the Debye-Waller factor is
\begin{equation}
    2M = \frac{12h^{2}sin^{2}\theta}{mk_{B}\Theta_{D}\lambda^{2}}\left[\frac{1}{4}+\frac{T^{2}}{\Theta_{D}^{2}}\int_{0}^{\Theta_{D}/T}\frac{\xi d \xi}{e^{\xi}-1}\right]
\label{eq:two},
\end{equation}
where $m$ is the atom mass, $\theta$ the
scattering angle, and $\lambda$ the x-ray wavelength. To evaluate the results, we 
approximated $m$ as the average mass, weighted according to form factors and 
atom concentrations, yielding $\Theta_{D}$ = 430 K. This is in good agreement with the 
(more precise) M\"{o}ssbauer  value of 413 K. However, below
$T_{f}$ there is an additional increase, corresponding to an enhancement of 6.5 $\%$ between 
35 and 10 K. 

The mean square displacement can be obtained from the 
x-ray diffraction intensities in a similar way to the M\"{o}ssbauer recoilless fraction, 
although the x-ray average includes
static as well as dynamic lattice displacements, and scattering from Al as well as Fe. 
The x-ray results give
$\sqrt{\langle x^{2}\rangle}$ = 0.014 nm at 35 K, approximately 3 times
larger than the value found
via M\"{o}ssbauer. Note that these averages assume that all atoms are vibrating in an
equivalent way; if instead only one Fe atom per cell were involved in the anomalous vibrational mode, for example, the result would instead be $\sqrt{\langle x^{2}\rangle}$ $\approx$ 0.13 nm for this atom
(depending upon the phase
factors involved in the particular reflections observed). A localized vibration involving only
a single Al yields instead 0.52 nm. These large values seem unlikely to be accommodated by the lattice, despite the 
presence of the vacancy site (Fig.~\ref{fig:figA}). Rietveld refinement \cite{chi05prb} 
yielded room-temperature thermal ellipsoids (assumed isotropic) with sizes
in the range 0.01$-$0.03 nm. While the sensitivity to such parameters
for a single atom is not high for such a complex stucture, these results make it 
seem unlikely that the anomalous vibrational mode is a localized mode. More likely,
the difference between the x-ray and M\"{o}ssbauer results can be ascribed to 
larger-amplitude motion for the lighter Al atom, combined with displacements that are
in part quasi-static. The results are consistent with vibrations due to magnetic
interactions among Fe atoms which are modulated by random magnetic
fluctuations. In the spin glass regime, these fluctuations freeze
out, reducing the associated atomic motions. This is the only instance of this type of
behavior of which we are aware.

Returning to the M\"{o}ssbauer center shifts of  Fig.~\ref{fig:figD}, the
decrease with $T$ is a second-order Doppler (SOD)
shift \cite{Dickson1}, due to the activation of phonon vibrations.
This shift is proportional to
$\Delta E$/$E_{o}$ = $-\langle v^{2}\rangle$/2$c^{2}$.
A fit according to the Debye approximation \cite{Dickson1} 
is shown by the dashed curve, giving $\Theta_{D}$ = 414 K,
very close to the value obtained from $f$. Since the SOD shift
contains $\langle v^{2}\rangle$ rather than
$\langle x^{2}\rangle$, the center shift is typically sensitive
only to higher-frequency vibrational modes. No change can be
seen in the center shift for the temperature range of spin-glass freezing,
which implies that the low-temperature change in
$f$ is due to activation of a low frequency mode. For comparison,
the spin exchange frequency was estimated from NMR \cite{chi05prb} 
to be about 10$^{12}$ Hz, which corresponds to the frequency of a 
relatively low-energy acoustic phonon. Therefore, the observed
results are consistent with the activation of a vibrational mode which is
coupled to the $T$-independent spin fluctuations in the paramagnetic regime. 

Anomalous low-temperature atomic motion is seen in another Fe-Al
system: Fe localized at an interstitial site in irradiated fcc Al
exhibits such behavior \cite{vog76prl,pet80prl,pau81jssc}, and a similar
low-temperature M\"{o}ssbauer response. 
However, in
the present case the freeze-out is coupled to a magnetic freezing
transition. The behavior is also somewhat different from that 
observed for the Y$_{2}$Mo$_{2}$O$_{7}$ pyrochlore \cite{Kere01PRL},
in which a structural distortion accompanies the freezing transition.
In FeAl$_{2}$ the magnetic fluctuations activate a soft mode, which
however does not completely soften so as to cause a structural transition
(as evidenced by the absence of change in x-ray diffraction).

The closely related quasicrystalline and approximant phase
aluminides provide a useful comparison. For example,
Al$_{65}$Cu$_{20}$Fe$_{15}$ is a stable icosahedral phase, while
Fe$_{4}$Al$_{13}$ is a decagonal approximant, however the range of 
Fe-Al metastable binary quasicrystals does not extend to Fe concentrations
as large as in FeAl$_{2}$ \cite{goldman93,huttunen04}.
Quasicrystals exhibit significantly enhanced diffusion and
plasticity, which has been attributed to the motion of phasons,
characterized by coordinated atomic jumps which individually cover
a fraction of a lattice constant. There have been direct
observations of phason motion \cite{edag00prl,Abe03nat}, and
of anomalous motion at surprisingly low temperatures, extending however to crystalline as well
as quasicrystalline phases \cite{dolinsek02}.
It would be useful to further characterize the soft mode evidenced in
FeAl$_{2}$ in order to gain further understanding of the lattice vibrational properties
of this class of materials.

To summarize, from M\"{o}ssbauer and x-ray measurements we have demonstrated
that FeAl$_{2}$ undergoes a change in vibrational behavior which accompanies
its spin-glass transition. This change takes the form of a soft mode which is activated
by the magnetic fluctuations in the material. This result is somewhat different
from the behavior previously observed in frustrated magnetic systems, however
it seems likely that similar behavior might also be present in other concentrated
spin glasses.

\begin{acknowledgments}
This work was supported by the Robert A. Welch Foundation, Grant
No. A-1526, by the National Science Foundation (DMR-0103455), and
by Young Teachers Program of MOE China (EYTP) and National Natural
Science Foundation of China (Grant No.50372005).
\end{acknowledgments}

\bibliography{FeAl2moss}

\begin{thebibliography}{26}
\expandafter\ifx\csname natexlab\endcsname\relax\def\natexlab#1{#1}\fi
\expandafter\ifx\csname bibnamefont\endcsname\relax
  \def\bibnamefont#1{#1}\fi
\expandafter\ifx\csname bibfnamefont\endcsname\relax
  \def\bibfnamefont#1{#1}\fi
\expandafter\ifx\csname citenamefont\endcsname\relax
  \def\citenamefont#1{#1}\fi
\expandafter\ifx\csname url\endcsname\relax
  \def\url#1{\texttt{#1}}\fi
\expandafter\ifx\csname urlprefix\endcsname\relax\def\urlprefix{URL }\fi
\providecommand{\bibinfo}[2]{#2}
\providecommand{\eprint}[2][]{\url{#2}}

\bibitem[{\citenamefont{Asamitsu et~al.}(1995)\citenamefont{Asamitsu, Moritomo,
  Tomioka, Arima, and Tokura}}]{Asam95nat}
\bibinfo{author}{\bibfnamefont{A.}~\bibnamefont{Asamitsu}},
  \bibinfo{author}{\bibfnamefont{Y.}~\bibnamefont{Moritomo}},
  \bibinfo{author}{\bibfnamefont{Y.}~\bibnamefont{Tomioka}},
  \bibinfo{author}{\bibfnamefont{T.}~\bibnamefont{Arima}}, \bibnamefont{and}
  \bibinfo{author}{\bibfnamefont{Y.}~\bibnamefont{Tokura}},
  \bibinfo{journal}{Nature (London)} \textbf{\bibinfo{volume}{373}},
  \bibinfo{pages}{407} (\bibinfo{year}{1995}).

\bibitem[{\citenamefont{Ramirez}(1997)}]{Rami97JPC}
\bibinfo{author}{\bibfnamefont{A.~P.} \bibnamefont{Ramirez}},
  \bibinfo{journal}{J. Phys.: Condens. Matter} \textbf{\bibinfo{volume}{9}},
  \bibinfo{pages}{8171} (\bibinfo{year}{1997}).

\bibitem[{\citenamefont{Renner et~al.}(2002)\citenamefont{Renner, Aeppli, Kim,
  Soh, and Cheong}}]{Renn02nat}
\bibinfo{author}{\bibfnamefont{C.}~\bibnamefont{Renner}},
  \bibinfo{author}{\bibfnamefont{G.}~\bibnamefont{Aeppli}},
  \bibinfo{author}{\bibfnamefont{B.~G.} \bibnamefont{Kim}},
  \bibinfo{author}{\bibfnamefont{Y.~A.} \bibnamefont{Soh}}, \bibnamefont{and}
  \bibinfo{author}{\bibfnamefont{S.~W.} \bibnamefont{Cheong}},
  \bibinfo{journal}{Nature (London)} \textbf{\bibinfo{volume}{416}},
  \bibinfo{pages}{518} (\bibinfo{year}{2002}).

\bibitem[{\citenamefont{Becker et~al.}(2002)\citenamefont{Becker, Streng, Luo,
  Moshnyaga, Damaschke, Shannon, and Samwer}}]{Beck02PRL}
\bibinfo{author}{\bibfnamefont{T.}~\bibnamefont{Becker}},
  \bibinfo{author}{\bibfnamefont{C.}~\bibnamefont{Streng}},
  \bibinfo{author}{\bibfnamefont{Y.}~\bibnamefont{Luo}},
  \bibinfo{author}{\bibfnamefont{V.}~\bibnamefont{Moshnyaga}},
  \bibinfo{author}{\bibfnamefont{B.}~\bibnamefont{Damaschke}},
  \bibinfo{author}{\bibfnamefont{N.}~\bibnamefont{Shannon}}, \bibnamefont{and}
  \bibinfo{author}{\bibfnamefont{K.}~\bibnamefont{Samwer}},
  \bibinfo{journal}{Phys. Rev. Lett.} \textbf{\bibinfo{volume}{89}},
  \bibinfo{pages}{237203} (\bibinfo{year}{2002}).

\bibitem[{\citenamefont{Mathur and Littlewood}(2003)}]{Math03pto}
\bibinfo{author}{\bibfnamefont{N.}~\bibnamefont{Mathur}} \bibnamefont{and}
  \bibinfo{author}{\bibfnamefont{P.}~\bibnamefont{Littlewood}},
  \bibinfo{journal}{Physics Today} \textbf{\bibinfo{volume}{68}},
  \bibinfo{pages}{25} (\bibinfo{year}{2003}).

\bibitem[{\citenamefont{Keren and Gardner}(2001)}]{Kere01PRL}
\bibinfo{author}{\bibfnamefont{A.}~\bibnamefont{Keren}} \bibnamefont{and}
  \bibinfo{author}{\bibfnamefont{J.~S.} \bibnamefont{Gardner}},
  \bibinfo{journal}{Phys. Rev. Lett.} \textbf{\bibinfo{volume}{87}},
  \bibinfo{pages}{177201} (\bibinfo{year}{2001}).

\bibitem[{\citenamefont{Corby and Black}(1973)}]{Corb73acr}
\bibinfo{author}{\bibfnamefont{R.~N.} \bibnamefont{Corby}} \bibnamefont{and}
  \bibinfo{author}{\bibfnamefont{P.~J.} \bibnamefont{Black}},
  \bibinfo{journal}{Acta Cryst.} \textbf{\bibinfo{volume}{29}},
  \bibinfo{pages}{2669} (\bibinfo{year}{1973}).

\bibitem[{\citenamefont{Kraj{\v c}\'i and Hafner}(2002)}]{Kraj02JPC}
\bibinfo{author}{\bibfnamefont{M.}~\bibnamefont{Kraj{\v c}\'i}}
  \bibnamefont{and} \bibinfo{author}{\bibfnamefont{J.}~\bibnamefont{Hafner}},
  \bibinfo{journal}{J. Phys.: Condens. Matter} \textbf{\bibinfo{volume}{14}},
  \bibinfo{pages}{5755} (\bibinfo{year}{2002}).

\bibitem[{\citenamefont{Lue and Kuo}(2003)}]{Lue03JPC}
\bibinfo{author}{\bibfnamefont{C.~S.} \bibnamefont{Lue}} \bibnamefont{and}
  \bibinfo{author}{\bibfnamefont{Y.-K.} \bibnamefont{Kuo}},
  \bibinfo{journal}{J. Phys.: Condens. Matter} \textbf{\bibinfo{volume}{15}},
  \bibinfo{pages}{877} (\bibinfo{year}{2003}).

\bibitem[{\citenamefont{Lue et~al.}(2001)\citenamefont{Lue, \"Oner, Naugle, and
  Ross}}]{Lue01PRB}
\bibinfo{author}{\bibfnamefont{C.~S.} \bibnamefont{Lue}},
  \bibinfo{author}{\bibfnamefont{Y.}~\bibnamefont{\"Oner}},
  \bibinfo{author}{\bibfnamefont{D.~G.} \bibnamefont{Naugle}},
  \bibnamefont{and} \bibinfo{author}{\bibfnamefont{J.~H.} \bibnamefont{Ross},
  \bibfnamefont{Jr.}}, \bibinfo{journal}{Phys. Rev. B}
  \textbf{\bibinfo{volume}{63}}, \bibinfo{pages}{184405}
  (\bibinfo{year}{2001}).

\bibitem[{\citenamefont{Chi et~al.}(2005)\citenamefont{Chi, Li, Vagizov,
  Goruganti, and Ross}}]{chi05prb}
\bibinfo{author}{\bibfnamefont{J.}~\bibnamefont{Chi}},
  \bibinfo{author}{\bibfnamefont{Y.}~\bibnamefont{Li}},
  \bibinfo{author}{\bibfnamefont{F.~G.} \bibnamefont{Vagizov}},
  \bibinfo{author}{\bibfnamefont{V.}~\bibnamefont{Goruganti}},
  \bibnamefont{and} \bibinfo{author}{\bibfnamefont{J.~H.} \bibnamefont{Ross},
  \bibfnamefont{Jr.}}, \bibinfo{journal}{Phys. Rev. B}
  \textbf{\bibinfo{volume}{71}}, \bibinfo{pages}{024431}
  (\bibinfo{year}{2005}).

\bibitem[{\citenamefont{Xu et~al.}(1993)\citenamefont{Xu, Xu, Yang, and
  Ma}}]{Xu93CSB}
\bibinfo{author}{\bibfnamefont{Z.~X.} \bibnamefont{Xu}},
  \bibinfo{author}{\bibfnamefont{Z.~D.} \bibnamefont{Xu}},
  \bibinfo{author}{\bibfnamefont{H.~S.} \bibnamefont{Yang}}, \bibnamefont{and}
  \bibinfo{author}{\bibfnamefont{R.~Z.} \bibnamefont{Ma}},
  \bibinfo{journal}{Chinese Science Bulletin} \textbf{\bibinfo{volume}{38}},
  \bibinfo{pages}{1767} (\bibinfo{year}{1993}).

\bibitem[{\citenamefont{Rancourt and Ping}(1991)}]{Rau91NIB}
\bibinfo{author}{\bibfnamefont{D.~G.} \bibnamefont{Rancourt}} \bibnamefont{and}
  \bibinfo{author}{\bibfnamefont{J.~Y.} \bibnamefont{Ping}},
  \bibinfo{journal}{Nucl. Instrum. Meth. B} \textbf{\bibinfo{volume}{58}},
  \bibinfo{pages}{85} (\bibinfo{year}{1991}).

\bibitem[{\citenamefont{V\'ertes et~al.}(1979)\citenamefont{V\'ertes, Korecz,
  and Burger}}]{vertes79}
\bibinfo{author}{\bibfnamefont{A.}~\bibnamefont{V\'ertes}},
  \bibinfo{author}{\bibfnamefont{L.}~\bibnamefont{Korecz}}, \bibnamefont{and}
  \bibinfo{author}{\bibfnamefont{K.}~\bibnamefont{Burger}},
  \emph{\bibinfo{title}{M\"{o}ssbauer Spectroscopy}}
  (\bibinfo{publisher}{Elsevier Scientific}, \bibinfo{year}{1979}),
  p.~\bibinfo{pages}{32}.

\bibitem[{\citenamefont{Vogl et~al.}(1976)\citenamefont{Vogl, Mansel, and
  Dederichs}}]{vog76prl}
\bibinfo{author}{\bibfnamefont{G.}~\bibnamefont{Vogl}},
  \bibinfo{author}{\bibfnamefont{W.}~\bibnamefont{Mansel}}, \bibnamefont{and}
  \bibinfo{author}{\bibfnamefont{P.~H.} \bibnamefont{Dederichs}},
  \bibinfo{journal}{Phys. Rev. Lett.} \textbf{\bibinfo{volume}{36}},
  \bibinfo{pages}{1497} (\bibinfo{year}{1976}).

\bibitem[{\citenamefont{Petry et~al.}(1980)\citenamefont{Petry, Vogl, and
  Mansel}}]{pet80prl}
\bibinfo{author}{\bibfnamefont{W.}~\bibnamefont{Petry}},
  \bibinfo{author}{\bibfnamefont{G.}~\bibnamefont{Vogl}}, \bibnamefont{and}
  \bibinfo{author}{\bibfnamefont{W.}~\bibnamefont{Mansel}},
  \bibinfo{journal}{Phys. Rev. Lett.} \textbf{\bibinfo{volume}{45}},
  \bibinfo{pages}{1862} (\bibinfo{year}{1980}).

\bibitem[{\citenamefont{Pauling}(1981)}]{pau81jssc}
\bibinfo{author}{\bibfnamefont{L.}~\bibnamefont{Pauling}}, \bibinfo{journal}{J.
  Solid State Chem.} \textbf{\bibinfo{volume}{40}}, \bibinfo{pages}{266}
  (\bibinfo{year}{1981}).

  \bibitem[{\citenamefont{Dattagupta}(1986)}]{Dickson1}
  \bibinfo{author}{\bibfnamefont{S.}~\bibnamefont{Dattagupta}}, in
    \emph{\bibinfo{booktitle}{M\"{o}ssbauer Spectroscopy}}, edited by
    \bibinfo{editor}{\bibfnamefont{D.~P.~E.} \bibnamefont{Dickson}}
    \bibnamefont{and} \bibinfo{editor}{\bibfnamefont{F.~J.} \bibnamefont{Berry}}
    (\bibinfo{publisher}{Cambridge University Press}, \bibinfo{year}{1986}), p.
    \bibinfo{pages}{198}; \bibinfo{author}{\bibfnamefont{E.~R.} \bibnamefont{Bauminger}}
    \bibnamefont{and} \bibinfo{author}{\bibfnamefont{I.}~\bibnamefont{Nowik}},
  \emph{ibid.}, p. \bibinfo{pages}{219}.


\bibitem[{\citenamefont{Israel et~al.}(1997)\citenamefont{Israel, Jacob,
  Soubeyroux, Fruchart, Pinto, and Melamud}}]{Isra97JAC}
\bibinfo{author}{\bibfnamefont{A.}~\bibnamefont{Israel}},
  \bibinfo{author}{\bibfnamefont{I.}~\bibnamefont{Jacob}},
  \bibinfo{author}{\bibfnamefont{J.~L.} \bibnamefont{Soubeyroux}},
  \bibinfo{author}{\bibfnamefont{D.}~\bibnamefont{Fruchart}},
  \bibinfo{author}{\bibfnamefont{H.}~\bibnamefont{Pinto}}, \bibnamefont{and}
  \bibinfo{author}{\bibfnamefont{M.}~\bibnamefont{Melamud}},
  \bibinfo{journal}{J. Alloys Comp.} \textbf{\bibinfo{volume}{253}},
  \bibinfo{pages}{265} (\bibinfo{year}{1997}).

\bibitem[{\citenamefont{Nath et~al.}(2002)\citenamefont{Nath, Klencs\'ar,
  Kuzmann, Homonnay, V\'ertes, Simopoulos, Devlin, Kallias, Ramirez, and
  Cava}}]{nat02prb}
\bibinfo{author}{\bibfnamefont{A.}~\bibnamefont{Nath}},
  \bibinfo{author}{\bibfnamefont{Z.}~\bibnamefont{Klencs\'ar}},
  \bibinfo{author}{\bibfnamefont{E.}~\bibnamefont{Kuzmann}},
  \bibinfo{author}{\bibfnamefont{Z.}~\bibnamefont{Homonnay}},
  \bibinfo{author}{\bibfnamefont{A.}~\bibnamefont{V\'ertes}},
  \bibinfo{author}{\bibfnamefont{A.}~\bibnamefont{Simopoulos}},
  \bibinfo{author}{\bibfnamefont{E.}~\bibnamefont{Devlin}},
  \bibinfo{author}{\bibfnamefont{G.}~\bibnamefont{Kallias}},
  \bibinfo{author}{\bibfnamefont{A.~P.} \bibnamefont{Ramirez}},
  \bibnamefont{and} \bibinfo{author}{\bibfnamefont{R.~J.} \bibnamefont{Cava}},
  \bibinfo{journal}{Phys. Rev. B} \textbf{\bibinfo{volume}{66}},
  \bibinfo{pages}{212401} (\bibinfo{year}{2002}).

\bibitem[{\citenamefont{Dong et~al.}(1999)\citenamefont{Dong, Wu, and
  Chen}}]{don99jac}
\bibinfo{author}{\bibfnamefont{C.}~\bibnamefont{Dong}},
  \bibinfo{author}{\bibfnamefont{F.}~\bibnamefont{Wu}}, \bibnamefont{and}
  \bibinfo{author}{\bibfnamefont{H.}~\bibnamefont{Chen}}, \bibinfo{journal}{J.
  Appl. Cryst.} \textbf{\bibinfo{volume}{32}}, \bibinfo{pages}{850}
  (\bibinfo{year}{1999}).

\bibitem[{\citenamefont{Goldman and Kelton}(1993)}]{goldman93}
\bibinfo{author}{\bibfnamefont{A.~I.} \bibnamefont{Goldman}} \bibnamefont{and}
  \bibinfo{author}{\bibfnamefont{R.~F.} \bibnamefont{Kelton}},
  \bibinfo{journal}{Rev. Mod. Phys} \textbf{\bibinfo{volume}{65}},
  \bibinfo{pages}{213} (\bibinfo{year}{1993}).

\bibitem[{\citenamefont{Huttunen-Saarivirta}(2004)}]{huttunen04}
\bibinfo{author}{\bibfnamefont{E.}~\bibnamefont{Huttunen-Saarivirta}},
  \bibinfo{journal}{J. Alloys Comp.} \textbf{\bibinfo{volume}{363}},
  \bibinfo{pages}{150} (\bibinfo{year}{2004}).

\bibitem[{\citenamefont{Edagawa et~al.}(2000)\citenamefont{Edagawa, Suzuki, and
  Takeuchi}}]{edag00prl}
\bibinfo{author}{\bibfnamefont{K.}~\bibnamefont{Edagawa}},
  \bibinfo{author}{\bibfnamefont{K.}~\bibnamefont{Suzuki}}, \bibnamefont{and}
  \bibinfo{author}{\bibfnamefont{S.}~\bibnamefont{Takeuchi}},
  \bibinfo{journal}{Phys. Rev. Lett.} \textbf{\bibinfo{volume}{85}},
  \bibinfo{pages}{1674} (\bibinfo{year}{2000}).

\bibitem[{\citenamefont{Abe et~al.}(2003)\citenamefont{Abe, Pennycook, , and
  Tsai}}]{Abe03nat}
\bibinfo{author}{\bibfnamefont{E.}~\bibnamefont{Abe}},
  \bibinfo{author}{\bibfnamefont{S.~J.} \bibnamefont{Pennycook}}, ,
  \bibnamefont{and} \bibinfo{author}{\bibfnamefont{A.~P.} \bibnamefont{Tsai}},
  \bibinfo{journal}{Nature (London)} \textbf{\bibinfo{volume}{421}},
  \bibinfo{pages}{347} (\bibinfo{year}{2003}).

\bibitem[{\citenamefont{Dolin{\v s}ek et~al.}(2002)\citenamefont{Dolin{\v s}ek,
  Apih, Jegli\u{c}, Feuerbacher, Calvo-Dahlborg, Dahlborg, and
  Dubois}}]{dolinsek02}
\bibinfo{author}{\bibfnamefont{J.}~\bibnamefont{Dolin{\v s}ek}},
  \bibinfo{author}{\bibfnamefont{T.}~\bibnamefont{Apih}},
  \bibinfo{author}{\bibfnamefont{P.}~\bibnamefont{Jegli\u{c}}},
  \bibinfo{author}{\bibfnamefont{M.}~\bibnamefont{Feuerbacher}},
  \bibinfo{author}{\bibfnamefont{M.}~\bibnamefont{Calvo-Dahlborg}},
  \bibinfo{author}{\bibfnamefont{U.}~\bibnamefont{Dahlborg}}, \bibnamefont{and}
  \bibinfo{author}{\bibfnamefont{J.~M.} \bibnamefont{Dubois}},
  \bibinfo{journal}{Phys. Rev. B} \textbf{\bibinfo{volume}{65}},
  \bibinfo{pages}{212203} (\bibinfo{year}{2002}).

\end{thebibliography}

\end{document}